\documentclass{article}
\usepackage{spconf,amsmath,amssymb,graphicx,hyperref}
\usepackage{booktabs,url,orcidlink}
\usepackage{microtype}
\hypersetup{hidelinks}

\title{Audio Preprocessing Effects on Stuttering Detection:\\
A Class-Specific Analysis}

\name{\begin{tabular}{@{}c@{}}Anisha Pattanayak\scalebox{0.75}{\orcidlink{0009-0005-2556-4472}}\textsuperscript{1},
Hanie Kang\scalebox{0.75}{\orcidlink{0009-0005-8023-9672}}\textsuperscript{2},
Huang-Cheng Chou\scalebox{0.75}{\orcidlink{0000-0003-2125-5689}}\textsuperscript{3},
Sudarsana Reddy Kadiri\scalebox{0.75}{\orcidlink{0000-0001-5806-3053}}\textsuperscript{3}\end{tabular}}
\address{\textsuperscript{1}Ming Hsieh Department of Electrical and Computer Engineering, University of Southern California, USA\\
\textsuperscript{2}Thomas Lord Department of Computer Science, University of Southern California, USA\\
\textsuperscript{3}Signal Analysis and Interpretation Laboratory (SAIL), University of Southern California, USA}

\begin{document}

\maketitle

\begin{abstract}
Audio preprocessing can affect how well a system detects stuttering.
We study a simulated chain of denoising, loudness normalisation, Opus
coding, and voice activity detection on SEP-28k. We use frozen WavLM
Base+ features and report pointwise confidence intervals from
episode-level bootstrap resampling. At a fixed threshold of $0.5$,
the chain reduces block $F_1$ from $0.638$ to $0.465$, with smaller
decreases for the other four classes. ROC-AUC decreases for all five
classes. Blocks show the largest $F_1$ and ROC-AUC losses, while sound
repetitions show the largest average precision loss. Tuning the threshold
on processed validation audio raises block $F_1$ to $0.630$. Retraining
on processed audio with threshold tuning gives $0.628$. Threshold
adjustment therefore accounts for most of the observed block $F_1$
recovery. It does not change ROC-AUC, which retraining raises only
from $0.620$ to $0.633$, compared with $0.724$ on clean audio.
\end{abstract}


\section{Introduction}
\label{sec:intro}

Automatic stuttering event detection is normally developed and evaluated on
corpus audio as distributed
\cite{lea2021sep28k,bayerl2022dysfluency,sheikh2023advancing,shih2024wordlevel}.
SEP-28k contains podcast recordings that may already have undergone
audio processing. We use the distributed recordings as our baseline,
referred to as clean audio, and apply additional simulated preprocessing.
We study operations that may occur before audio reaches a deployed
detector: noise suppression, automatic gain control, voice activity
detection (VAD) \cite{sohn1999vad}, and speech coding. These stages act
differently. A noise suppressor and a VAD are built to attenuate or discard
what they classify as non-speech. A codec allocates bits under a model of
speech and an automatic gain control rescales the signal, so neither
discards non-speech by design, although both can still alter low-energy
regions.

This may matter for stuttering because one dysfluency type is defined by the
absence of sound. In a silent block the speaker is stuck, and the acoustic
trace is silence or near-silence that a VAD can label as non-speech. Other
types leave audible energy: a prolongation is sustained voicing, and an
interjection is a filler word that is usually voiced, so a silence remover
is more likely to keep them. Whether the effect differs by class is an
empirical question, and the direction is not obvious, because removing
frames also changes the context around an event. We train the detector on
clean audio, test it on audio passed through a simulated chain, and ask how
much the chain costs the detector class by class and whether that depends on
the metric; which stage accounts for most of it, and how VAD-targeted
deletion compares with duration-matched random deletion; and how much is
recovered by adapting the decision threshold alone, and how much more by
retraining on matched audio. Code and results are
released.\footnote{\small\href{https://github.com/anishxa/preprocessing-effects-on-stuttering}{\nolinkurl{github.com/preprocessing-effects-on-stuttering}}}

\section{Related work}
\label{sec:related}

Endpointing and VAD are known to disadvantage people who stutter. Lea et
al.\ \cite{lea2023chi} surveyed 61 people who stutter and analysed
recordings from 91 participants; retuning the endpointing cut premature
termination by $79.1\%$ and word error rate from $25.4\%$ to $9.9\%$. That
work concerns recognition and user experience, not what a front-end does to
a dysfluency \emph{detector}.

Work on stuttering event detection has instead studied representations:
self-supervised encoders \cite{bayerl2022dysfluency,shih2024wordlevel} and
augmentation with class-balanced losses \cite{sheikh2023advancing}. Data
partitioning has been examined \cite{bayerl2022partitioning}, and a scoping
review notes that deployment guidance lags behind model building
\cite{scoping2026}. We are not aware of a study that measures the per-class
effect of a preprocessing chain on this task, decomposes it stage by stage,
and separates what a tuned threshold recovers from what retraining
recovers.

\section{Experimental Setup}
\label{sec:method}

\subsection{Dataset and labels}
\label{sec:data}
 
SEP-28k \cite{lea2021sep28k} gives 3~s podcast clips labelled by three
listeners for five dysfluency types. Each label is a count
$c \in \{0,1,2,3\}$ of listeners who marked the event, and a clip is positive
for class $k$ when $c_k \ge 1$. That threshold keeps events only one of the
three listeners heard, so the positive sets include disputed labels. After
dropping clips marked as music or as containing no speech, a fixed subset of
$N = 8{,}000$ clips is drawn with seed 42: the sampler takes up to 1{,}300
clips carrying each class label, in the order word repetition, sound
repetition, prolongation, block, interjection, assigning each clip once, then
fills the remainder with clips carrying no event label. Because labels
overlap, the result is not class balanced: 3{,}670 blocks, 3{,}339
interjections, 2{,}704 prolongations, 2{,}337 sound repetitions and 2{,}000
word repetitions. In all, 6{,}500 clips ($81.2\%$) carry at least one label,
2.16 on average, over 241 episodes of five shows.
 
\subsection{Clip representation}
 
We take frozen features from WavLM Base+ \cite{chen2022wavlm} at 16~kHz. For
a clip with $T$ frames, hidden states $\mathbf{h}_1,\dots,\mathbf{h}_T$ and a
mask $m_t$ marking real frames, the clip embedding is the masked mean
 
\begin{equation}
\mathbf{z} \;=\; \frac{\sum_{t=1}^{T} m_t \, \mathbf{h}_t}
                      {\max\left(\sum_{t=1}^{T} m_t,\, 1\right)} .
\label{eq:pool}
\end{equation}
 
The mask matters because the chain changes clip length, and pooling over
padding would make $\mathbf{z}$ depend on how much audio was removed.
 
\subsection{Processing conditions}
 
We build ten processed versions of every clip, plus the clean original, all
simulated with \texttt{ffmpeg}, \texttt{webrtcvad} and spectral gating
rather than captured from a live audio processing module. Three are Opus
\cite{valin2012opus} at 16 and 8~kbit/s and at 16~kbit/s in VoIP mode; we
make no claim about discontinuous transmission. One is spectral-gating
denoising. Two apply WebRTC VAD at aggressiveness 3, either deleting the
non-speech frames and concatenating the rest, so the clip shortens, or
zeroing them in place. Two are chains: denoising, loudness normalisation to
$-23$~LUFS \cite{itu2015bs1770} and Opus at 16~kbit/s, without and with VAD
deletion. The last two delete randomly chosen frames, either matched per
clip to the duration the VAD condition removes or a random $30\%$, holding
the removed duration fixed while changing which frames are chosen. Clips
shorter than 400~ms are padded to that length, which affects 20 of 8,000
clips, or $0.25\%$, in the worst condition. The classifier is trained on
clean audio and tested on processed audio, the deployment case.
 
\subsection{Silence and duration statistics}
 
For a clean signal $x$ and its processed version $y$ we take root-mean-square
energies $r_i(\cdot)$ over 20~ms frames and set one silence threshold on the
clean signal: $\tau$ is the larger of $0.01$ and the 30th percentile of
$\{r_i(x)\}$. The same $\tau$ is applied to both signals. Writing
$S(\cdot)$ for the total duration of frames below $\tau$ and $D(\cdot)$ for
total duration,
 
\begin{equation}
\rho \;=\; 1 - \frac{S(y)}{S(x)}, \qquad
\delta \;=\; 1 - \frac{D(y)}{D(x)} ,
\label{eq:silence}
\end{equation}
 
give the fraction of silence removed and the fraction of duration lost. Both
are clipped to $[0,1]$, computed per clip before padding, and describe the
signal rather than a mediator in a causal model.
 
\subsection{Threshold selection}
\label{sec:thresh}
 
Every tuned threshold reported here comes from one nested procedure, run
separately in each outer fold. Inside the outer training set, one of the four
remaining episode-grouped folds is held out as an inner validation split and
the other three form the inner training set. An inner probe is fitted on the
inner training set and applied to the inner validation split, and for each
class independently the threshold is chosen from 91 values evenly spaced on
$[0.05, 0.95]$ to maximise $F_1$ on that split. For the clean baseline, the inner classifier is trained and validated
on clean features. For threshold-only adaptation, the same clean-trained
classifier is evaluated on processed validation features to select the
threshold. For matched retraining, the inner classifier is trained on
processed training features and evaluated on processed validation
features. The
selected per-class thresholds are then applied to the outer model, which is
refitted on the whole outer training set, and evaluated once on the held-out
outer test fold. That test fold enters neither layer selection nor threshold
search, so no test label is used in either choice.
 
\begin{figure*}[t]
\centering
\includegraphics[width=\textwidth]{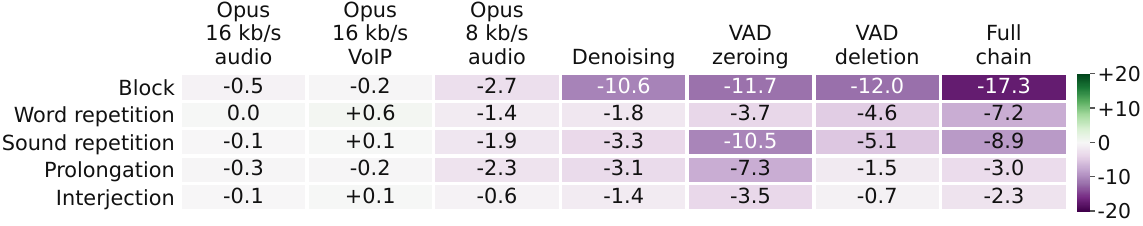}
\caption{Changes in pooled out-of-fold $F_1$ relative to clean audio, in
percentage points. All conditions use clean-trained classifiers with a
decision threshold of $0.5$. Positive values indicate improvement. Component
effects need not sum to the full-chain effect. Colors and annotations
indicate effect sizes, not statistical significance.}
\label{fig:f1_heatmap}
\end{figure*}
 
\subsection{Splits and uncertainty}
\label{sec:setup}
 
Evaluation uses five-fold cross-validation with folds grouped by episode, so
no episode appears in both training and test \cite{bayerl2022partitioning}.
Uncertainty is reported as $95\%$ confidence intervals from a paired cluster
bootstrap over episodes with $B = 1000$ resamples
on pooled out-of-fold predictions, since clips from one episode are not
independent. Each resample draws episodes with replacement and scores the
clean and processed versions of the same clips, so an interval on a
difference is paired. These intervals are pointwise: they are not adjusted
for the number of classes or conditions, and one interval excluding zero is
not a multiplicity-controlled test. We report no hypothesis test for any
condition comparison.
 
\subsection{Layer selection, classifier and metrics}
 
The encoder layer is chosen by an inner cross-validation inside each outer
training fold, so no test episode influences the choice; the selected layers
were 10, 8, 9, 9 and 8. The classifier is one-versus-rest logistic
regression with balanced class weights. We report
$F_1$ at a fixed threshold of $0.5$, the area under the ROC curve (AUC), and
average precision (AP), computed with \texttt{average\_precision\_score}.
More than one metric matters because $F_1$ at a fixed threshold mixes a
change in ranking quality with a change in the best operating point, while
AUC and AP are invariant to the second.
 
\section{Results}
\label{sec:results}
 
\subsection{The effect depends on the class and on the metric}
 
Figure~\ref{fig:f1_heatmap} summarises the class-specific changes at the
fixed threshold, with purple for a decrease and green for an increase. The
full chain produces the largest observed $F_1$ decrease for blocks, $17.3$
points, followed by sound repetitions at $8.9$ and word repetitions at $7.2$.
VAD zeroing and VAD deletion have similar effects for blocks but differ for
the other classes.
 
Table~\ref{tab:main} adds intervals. The block loss is $27.1\%$ relative,
all five intervals on $\Delta F_1$ lie below zero, and all five classes lose
AUC. The metrics do not rank the classes identically: blocks show the
largest $F_1$ and AUC decrease, but sound repetitions show the largest AP
decrease, $10.5$ points against $8.9$, while word repetitions lose $7.2$
points of $F_1$ and only $1.8$ of AP.
 
\begin{table}[t]
\centering
\caption{Clean audio versus the full chain, per class, at threshold $0.5$.
Changes are in percentage points, with pointwise $95\%$ bootstrap intervals
on $\Delta F_1$. AP is average precision.}
\label{tab:main}
\small
\setlength{\tabcolsep}{2.5pt}
\begin{tabular}{@{}lrcrr@{}}
\toprule
Class & clean & $\Delta F_1$ [95\% CI] & $\Delta$AUC & $\Delta$AP \\
\midrule
Block        & 0.638 & $-17.3$ $[-19.5,-15.2]$ & $-10.4$ & $-8.9$ \\
SoundRep     & 0.659 & $-8.9$ $[-10.6,-7.3]$ & $-8.3$ & $-10.5$ \\
WordRep      & 0.645 & $-7.2$ $[-9.0,-5.5]$ & $-2.8$ & $-1.8$ \\
Prolong.     & 0.624 & $-3.0$ $[-4.3,-1.7]$ & $-3.2$ & $-4.1$ \\
Interj.      & 0.756 & $-2.3$ $[-3.1,-1.3]$ & $-1.9$ & $-2.1$ \\
\bottomrule
\end{tabular}
\end{table}
 
\subsection{Stage decomposition and the deletion controls}
\label{sec:stages}
 
Table~\ref{tab:cond} decomposes the chain for block detection. Opus changes
block $F_1$ by at most $2.7$ points, at 8~kbit/s. At threshold $0.5$,
denoising and VAD deletion produce similar block $F_1$ decreases. After
validation-based threshold tuning, the decreases relative to tuned clean
performance are approximately $1.6$ and $3.6$ points, and their ROC-AUC
decreases are approximately $2.8$ and $7.2$ points, so denoising is the milder
of the two once the operating point is adapted. The chain with VAD gating is
$7.4$ points lower at threshold $0.5$ than the same chain without it.
By Eq.~\eqref{eq:silence}, VAD deletion removes $78\%$ of the clean silence
and $41\%$ of the duration, the full chain $72\%$ and $47\%$, and matched
random deletion $42\%$ and $41\%$; the codec and denoising conditions remove
neither.
 
At the fixed threshold, deleting randomly chosen frames, matched per clip to
the duration that VAD gating removes, changes block $F_1$ by $1.1$ points
with an interval that includes zero, while VAD-targeted deletion changes it
by $12.0$ points, although the removed duration is the same. With thresholds
tuned on processed validation data, block $F_1$ is $0.632$ for VAD deletion,
$0.631$ for VAD zeroing and $0.647$ for matched random deletion, against a
tuned clean baseline of $0.668$, so the gap shrinks from $0.109$ to $0.015$,
while ROC-AUC falls in all three, from $0.724$ to $0.653$, $0.650$ and
$0.662$. Ranking quality degrades whether the removed frames are chosen by a
VAD or at random.
 
The class pattern also changes with the threshold. At threshold $0.5$,
matched random deletion is the worst of the three deletion conditions for
prolongations and nearly harmless for blocks, while VAD deletion is the
reverse. After tuning it is the worst for four of the five classes: against
each class's tuned clean baseline, the tuned decreases for matched random
deletion against VAD deletion are $8.6$ and $3.7$ points for sound
repetitions, $10.3$ and $0.8$ for word repetitions, $6.2$ and $0.4$ for
prolongations, and $6.8$ and $1.0$ for interjections, with blocks the
exception at $2.1$ against $3.6$. These are per-condition comparisons
that Table~\ref{tab:adapt}, covering the full chain only, does not verify.
We therefore do not claim that the loss follows which frames are removed
rather than how many.
 
Deletion and zeroing share their VAD decisions and differ only in whether
rejected frames are removed or silenced in place. For blocks the two are
close, $0.518$ against $0.520$ in pooled $F_1$; for other classes zeroing is
worse, with sound repetitions losing $10.5$ points against $5.1$. Close point
estimates do not establish equivalence and do not rule out a contribution
from shortening.
 
\begin{table}[t]
\centering
\caption{Block detection by condition, with clean-trained classifiers, in
percentage points. The \emph{fix} column uses threshold $0.5$ and carries
pointwise $95\%$ bootstrap intervals; \emph{tun} tunes the threshold as in
Section~\ref{sec:thresh} and has no interval.}
\label{tab:cond}
\small
\setlength{\tabcolsep}{2.5pt}
\begin{tabular}{@{}lcrr@{}}
\toprule
Condition & $\Delta F_1$ fix [95\% CI] & $\Delta F_1$ tun & $\Delta$AUC \\
\midrule
Opus 16k       & $-0.53$ $[-1.05,-0.05]$ & $+0.08$ & $-0.17$ \\
Opus 16k VoIP  & $-0.22$ $[-0.67,+0.20]$ & $+0.07$ & $-0.14$ \\
Opus 8k        & $-2.73$ $[-3.59,-1.90]$ & $-0.82$ & $-1.00$ \\
Denoise        & $-10.59$ $[-12.59,-8.82]$ & $-1.64$ & $-2.78$ \\
VAD zero       & $-11.73$ $[-13.51,-10.09]$ & $-3.73$ & $-7.46$ \\
VAD delete     & $-11.98$ $[-13.70,-10.31]$ & $-3.58$ & $-7.16$ \\
Chain, no VAD  & $-9.91$ $[-11.66,-8.22]$ & $-1.21$ & $-2.71$ \\
Full chain     & $-17.31$ $[-19.45,-15.23]$ & $-3.82$ & $-10.43$ \\
Random matched & $-1.09$ $[-2.22,+0.13]$ & $-2.12$ & $-6.19$ \\
Random $30\%$  & $-0.78$ $[-1.71,+0.19]$ & $-1.54$ & $-2.68$ \\
\bottomrule
\end{tabular}
\end{table}
 
\subsection{Threshold adaptation and matched retraining}
 
Table~\ref{tab:adapt} separates the two adaptations under the full chain.
Tuning the decision threshold on processed validation data, with the
clean-trained model unchanged, raises block $F_1$ from $0.465$ to $0.630$.
Training on matched processed audio and then tuning gives $0.628$. Against
the fixed-threshold clean-to-degraded gap, threshold tuning alone closes
$95.3\%$ and matched retraining with tuning closes $94.2\%$. Threshold
adaptation therefore accounts for most of the observed block $F_1$ recovery,
and matched retraining does not improve the block point estimate beyond it.
We ran no paired comparison of the two adapted models, so we do not claim
that they are equivalent or that either is better. Other classes differ:
against threshold tuning alone, matched retraining adds $1.6$ points for
sound repetitions, $2.9$ for word repetitions and $0.6$ for interjections,
and subtracts $1.0$ for prolongations.
 
A threshold change cannot alter ROC-AUC, so the AUC columns compare the
clean-trained and matched-trained models only. Matched training raises block
AUC from $0.620$ to $0.633$, a modest improvement that leaves it far below
the clean value of $0.724$. The same holds elsewhere, from $0.761$ to $0.789$
for sound repetitions against $0.844$ clean, while prolongations do not
improve. Matched training therefore recovers part of the lost ranking
quality and not all of it.
 
\begin{table}[t]
\centering
\caption{Threshold adaptation against matched retraining, full chain only.
$F_1$: \emph{clean-t} is clean audio at a clean-tuned threshold, \emph{thr}
is processed audio with the clean-trained model at a processed-tuned
threshold, and \emph{mat} adds matched retraining. AUC columns are clean,
processed clean-trained and processed matched-trained; AUC does not depend
on the threshold. Point estimates, no intervals.}
\label{tab:adapt}
\small
\setlength{\tabcolsep}{4pt}
\begin{tabular}{@{}lrrrrrr@{}}
\toprule
 & \multicolumn{3}{c}{$F_1$} & \multicolumn{3}{c}{AUC} \\
\cmidrule(lr){2-4}\cmidrule(lr){5-7}
Class & clean-t & thr & mat & clean & proc & mat \\
\midrule
Block        & 0.668 & 0.630 & 0.628 & 0.724 & 0.620 & 0.633 \\
SoundRep     & 0.657 & 0.579 & 0.595 & 0.844 & 0.761 & 0.789 \\
WordRep      & 0.643 & 0.607 & 0.636 & 0.851 & 0.823 & 0.841 \\
Prolong.     & 0.618 & 0.604 & 0.594 & 0.789 & 0.757 & 0.755 \\
Interj.      & 0.757 & 0.734 & 0.740 & 0.866 & 0.846 & 0.849 \\
\bottomrule
\end{tabular}
\end{table}
 
\subsection{Sensitivity checks}
\label{sec:sens}
 
With majority-vote labels ($c_k \ge 2$) every class scores lower on clean
audio, most for blocks ($0.637$ to $0.346$). This check relabels the
evaluation targets only: it reuses the same clean-trained model and
threshold of $0.5$ and does not retrain or retune, so it bounds
label-threshold sensitivity rather than reporting a majority-vote system.
Replacing the linear probe with a small MLP does not help, at $0.014$ lower
for blocks on clean audio and $0.082$ lower under the chain. Calibration
error rises under the chain for blocks, from $0.036$ to $0.095$, which again
points to an operating-point effect. Training on a subset of shows and
testing on held-out shows reproduces the ordering, with block $F_1$ falling
from $0.667$ to $0.453$; there recall falls from $0.718$ to $0.383$ while
precision falls only from $0.622$ to $0.555$, so the detector mostly stops
finding blocks rather than inventing them. Aggregating predictions to
episodes with a weighted event-rate proxy, the chain lowers that proxy by
$9.4\%$ $[-12.3, -6.2]$ against the clean-audio estimate, while against
label-derived severity the change includes zero. That proxy is
model-derived, not a validated clinical measure.
 
\section{Limitations}
\label{sec:limits}
 
The chain is simulated with \texttt{ffmpeg}, \texttt{webrtcvad} and spectral
gating rather than a live audio processing module, so magnitudes may differ
from a specific product. The comparisons are observational contrasts on the
same clips, not a mediation analysis, and $\rho$ and $\delta$ are
descriptive. The loss under zeroing may partly reflect exact digital zeros,
which the encoder never meets in training. The tuned columns in
Table~\ref{tab:cond} and all of Table~\ref{tab:adapt} are point estimates
without intervals. Results come from one corpus of five shows, and the
cross-show test is a shift within it, not external replication. The frozen
encoder and linear probe make absolute scores uncompetitive by design, so our
claims concern relative change, and evaluation is clip-level on fixed 3~s
windows rather than temporal event detection \cite{scoping2026}.
 
\section{Conclusion}
\label{sec:conclusion}
 
Simulated preprocessing lowers stuttering event detection unevenly across
dysfluency types, most for silent blocks, and all five classes lose ROC-AUC.
Tuning the decision threshold on processed data restores block $F_1$ to
$0.630$ while matched retraining with tuning gives $0.628$, so threshold
adaptation accounts for most of the recovery, while matched training lifts
block ROC-AUC only from $0.620$ to $0.633$ against $0.724$ clean. Reporting
$F_1$ at one fixed threshold confounds lost ranking quality with a shifted
operating point.
 
\newpage
\bibliographystyle{IEEEbib}
\bibliography{refs}
 
\end{document}